\documentclass{IEEEtran}
\usepackage{cite}
\usepackage{amsmath,amssymb,amsfonts}
\usepackage{graphicx}
\usepackage{textcomp,nicefrac}
\usepackage{romannum}
\usepackage{array}
\usepackage{float}
\graphicspath{ {./plots/} }
\def\BibTeX{{\rm B\kern-.05em{\sc i\kern-.025em b}\kern-.08em
T\kern-.1667em\lower.7ex\hbox{E}\kern-.125emX}}
\begin{document}
\title{The Analog Front-end for the LGAD Based Precision Timing Application in CMS ETL}
\author{Quan Sun, Sunil M. Dogra, Christopher Edwards, Datao Gong, Lindsey Gray, Xing Huang, Siddhartha Joshi , Jongho Lee, Chonghan Liu, Tiehui Liu, Tiankuan Liu, Sergey Los, Chang-Seong Moon, Geonhee Oh, Jamieson Olsen, Luciano Ristori, Hanhan Sun, Xiao Wang, Jinyuan Wu, Jingbo Ye, Zhenyu Ye, Li Zhang and Wei Zhang
\thanks{This manuscript has been authored by Fermi Research Alliance, LLC under Contract No. DE-AC02-07CH11359 with the U.S. Department of Energy, Office of Science, Office of High Energy Physics.}
\thanks{Quan Sun, Christopher Edwards, Lindsey Gray, Sergey Los, Jamieson Olsen, Luciano Ristori, Jinyuan Wu and Tiehui Liu are with Fermi National Accelerator Laboratory, Batavia, Illinois, USA. (Corresponding authors: Quan Sun and Tiehui Ted Liu, e-mail: qsun@fnal.gov and thliu@fnal.gov).}
\thanks{Datao Gong, Chonghan Liu, Tiankuan Liu and Jingbo Ye are with Southern Methodist University, Dallas, Texas, USA.}
\thanks{Sunil Dogra, Chang-Seong Moon and Jongho Lee are with Kyungpook National University, Daegu, South Korea. The work of them has been supported by the National Research Foundation of Korea (NRF) grant funded by the Korea government (MSIT) (Grants No. 2018R1A6A1A06024970, No. 2020R1A2C1012322 and Contract NRF-2008-00460).}
\thanks{Siddhartha Joshi was with Northwestern University, Evanston Illinois, USA. He is now with Intel Corporation, Hillsboro, Oregon, USA.}
\thanks{Geonhee Oh, Xiao Wang and Zhenyu Ye are with University of Illinois at Chicago, Chicago, IL, USA.}
\thanks{Xing Huang, Hanhan Sun, Li Zhang and Wei Zhang are with Southern Methodist University, Dallas, Texas, USA and Central China Normal University, Wuhan, Hubei, P.R. China}
}

\maketitle

\begin{abstract}
The analog front-end for the Low Gain Avalanche Detector (LGAD) based
precision timing application in the CMS Endcap Timing Layer (ETL) has been
prototyped in a 65 nm CMOS mini-ASIC named ETROC0. Serving as the very first
prototype of ETL readout chip (ETROC), ETROC0 aims to study and demonstrate
the performance of the analog front-end, with the goal to achieve 40 to 50
ps time resolution per hit with LGAD (therefore reach about 30ps per track
with two detector-layer hits per track). ETROC0 consists of preamplifier and
discriminator stages, which amplifies the LGAD signal and generates digital
pulses containing time of arrival and time over threshold
information. This paper will focus on the design considerations that lead to the ETROC front-end architecture choice, the key design features of the building blocks, the methodology of using the LGAD simulation data to evaluate and optimize the front-end design. The ETROC0 prototype chips have been extensively tested using charge injection and the measured performance agrees well with simulation. The initial beam test results are also presented, with time resolution of around 33 ps observed from the preamplifier waveform analysis and around 41 ps from the discriminator pulses analysis.  A subset of ETROC0 chips have also been tested to a total ionizing dose of 100 MRad with X-ray and no performance degradation been observed.  
\end{abstract}


\section{Introduction}
\label{sec:introduction}
\IEEEPARstart{T}{he} new MIP Timing Detector (MTD)\cite{CMS:2667167} has
been officially approved by the Compact Muon Solenoid (CMS) experiment for
the High-Luminosity Large Hadron Collider (HL-LHC) upgrade. It is
aiming to measure the arrival time of charged particles with a time
resolution of 30 to 40 ps per track. The MTD consists of barrel and endcap
sub-detectors. The endcap sub-detector or the endcap timing layer (ETL) is based on Low Gain Avalanche
Detector (LGAD) \cite{White_2014,PELLEGRINI201412,CARTIGLIA2015141}. It is designed to have two-layer hits for a given track such that the required time resolution per hit is in the range between 40 to 50 ps.

Besides the stringent precision time measurement requirement, the radiation
level pose additional challenge on ETL LGAD sensors. A large fraction of ETL LGAD sensors will receive
a rather mild dose: 50\% of the sensors will be exposed to a fluence of less
than $5\times10^{14} n_{eq}/cm^2$, 80\% to less than $8\times10^{14}
n_{eq}/cm^2$, while the innermost 10\% of the detector will be exposed to
more than $1\times10^{15}n_{eq}/cm^2$. The total ionizing dose (TID)
accumulated on the readout electronics is foreseen to be 100 MRad in the
worst case.

The ETL readout ASIC chip (ETROC) will be designed to handle a 16×16 pixel cell
matrix, each pixel cell being 1.3×1.3 $mm^2$ to match with the LGAD sensor
pixel size. At the pixel cell level, each channel consists of a
preamplifier, a discriminator, a TDC used to digitize the time of
arrival (TOA) and time over threshold(TOT) measurements, and a memory for data
storage and readout logics. The TOT is used for time-walk correction of the TOA
measurement. The design goal for the time resolution is 50 ps per hit, in
order to achieve a 35 ps arrival time measurement for a track with a hit in each of the two layers. Towards the end of HL-LHC operation, the
time resolution should be kept below 60 ps per track as the LGAD gain drops
with irradiation. The time resolution is due to a combination of sensor and readout electronics performance. The LGAD contribution is known to be
about 30 ps at the beginning of operation, and is expected to moderately
degrade to 40 ps towards the end of HL-LHC operation for the innermost 10\%
of the detector. The
electronics contribution is expected to be dominated by
preamplifier/discriminator stage, and the contribution from the
preamplifier/discriminator is designed to be below 40 ps, which should be
achieved with reasonable power consumption and signal efficiency, and
maintained throughout the HL-LHC operation. The power consumption of ETROC is constrained by the cooling system of ETL and should
be within 1 W/chip. This means that, after excluding the common circuitry, the power consumption of each channel or pixel should be kept below 3 mW.

The analog front-end, including the preamplifier, the discriminator and the
threshold voltage generator, is critical for the CMS ETL precision timing
performance. In this paper, we present the analog front-end, which has been implemented in a prototype chip named ETROC0 in 65 nm CMOS technology. ETROC0 is the very first prototype and is designed and optimized for LGAD sensors with size of 1.3 x 1.3 \(mm^2\) and the effective thickness of 50 $\mu$m, to achieve a time resolution better than 50 ps per hit while consuming very low power. The second prototype chip is ETROC1 which includes a 4 x 4 active pixel matrix with a precision clock distribution network. In each pixel, the same analog front-end from ETROC0 is connected with a new TDC\cite{Wei_TDC2020} to form a full precision timing signal processing chain. The details of ETROC1 is beyond the scope of this paper.

In this paper, we present the design and test results of ETROC0. 
Section \Romannum{2} discusses the design considerations of the analog front-end for ETL. 
ETROC0 circuit level design is presented in section \Romannum{3}. This is followed by testing results in section \Romannum{4} and a summary in section \Romannum{5}.

\section{Design consideration}
CMS ETL timing detector chooses LGAD sensors with the size of 1.3 x 1.3 \(mm^2\) and effective thickness of the 50 $\mu$m to achieve optimized time resolution, granularity and overall readout electronics power consumption trade-off\cite{CMS:2667167}. The front-end electronics expect about 15 fC most probable value (MPV) charge at a typical sensor bias voltage with which a gain of 20 is achieved in the sensor. The sensors will be operated at a relatively low biased voltage at the beginning of the operation, but a higher bias voltage is required to maintain a reasonable high enough gain towards the end of operation due to accumulated radiation damage. The capacitance and the rise time seen by the front-end electronics are 3.4 pF and 500 ps, respectively, which are key parameters used to optimize the front-end amplifier. The intrinsic time resolution of the LGAD is 30 ps, which set the lower limit of the time measurement system.

With the charge from a LGAD sensor converted into voltage in-situ over the sensor capacitance, a voltage step of only about 4.4 mV is seen by the front-end electronics. A preamplifier is needed to convert the charge into voltage pulses with an adequate gain before the timing discrimination. A discriminator takes the output of the preamplifier and makes the timing decision at a user defined threshold. There are two aspects to consider for the time resolution, one is the jitter and the other is the time walk. 

The jitter of the signal at the output of the amplifier could be expressed as:
\begin{equation}
jitter=\frac{v_n}{\frac{dV}{dt}}
\end{equation}
where $v_n$ is the output referred noise of the preamplifier and $\frac{dV}{dt}$ is the slope of the signal at the threshold crossing point where the timing decision is made by the discriminator. For the output referred noise and the slope,
\begin{equation}
v_n \varpropto \sqrt{f_u} \varpropto \frac{1}{\sqrt{t_r}}
\end{equation}
and
\begin{equation}
\frac{dV}{dt} \varpropto \frac{1}{t_r} \varpropto f_u
\end{equation}
are valid, where $f_u$ is the bandwidth of the preamplifier and $t_r$ is the rise time of the preamplifier output signal. When an LGAD sensor is as the stimulus of the preamplifier, the rise time becomes
\begin{equation}
t_c=\sqrt{t_s^2+t_r^2}
\end{equation}
, where $t_s$ is the rise time of the LGAD signal, and $t_c$ is the cumulative rise time. The slope of the output signal can be expressed as
\begin{equation}
\frac{dV}{dt} \approx \frac{v_o}{t_c},
\end{equation}
where $v_o$ is the amplitude of the output signal.
The timing jitter of the preamplifier with LGAD input can be expressed as
\begin{equation}
jitter=\frac{v_n}{\frac{dV}{dt}} \varpropto \frac{1}{\sqrt{t_r}}\cdot\frac{t_c}{v_o} \varpropto \frac{1}{v_o}\cdot\sqrt{t_s}\cdot\sqrt{\frac{t_s}{t_r}+\frac{t_r}{t_s}}.
\label{eqe:6}
\end{equation}
Equation \eqref{eqe:6} suggests for a given $t_s$, jitter is minimized when $t_r$ equals to $t_s$. However, the dependence is not strong and an approximate matching is
adequate. For example, a mismatch with a factor of two between $t_s$ and $t_r$ only increases the jitter by no more than 12\%. The optimum timing jitter improves with the decrease of $t_s$. Increasing output signal amplitude also helps to improve the optimum jitter. In a front-end including a preamplifier and a discriminator, the jitter from the discriminator is usually small with an appropriate design, making the dominant jitter contributor the preamplifier.

The jitter discussed so far is the noise-related timing variation at a given signal amplitude. In addition, the time at which the preamplifier signal crosses the threshold depends on the amplitude of the signal. As the amplitude varies, the time crossing point shifts. This way, the variation of the amplitude broadens the time variation. This is the time walk, another contributor to the time variation in time measurement. For CMS ETL, the energy deposited by charged particles on the LGAD varies according to Landau distribution\cite{Sadrozinski_2017,Meroli_2011}. The time walk is not negligible as the amplitude varies significantly.

With time walk presenting, the best time resolution could be achieved with waveform sampling and sophisticated digital signal processing\cite{GENAT2009387}. However, a reasonable waveform sampling implementation requires significant power consumption and data transmission bandwidth, which is not feasible for ETROC. 

The time variation introduced by time walk could be largely reduced with constant fraction discrimination (CFD) technique. In CFD implementations, a delayed version and an attenuated version of the original signal are compared in a comparator to make the time determination. This technique significantly removes the dependence of time variation on both the amplitude and the pulse shape\cite{spieler2005semiconductor,603660cfd}. However, the high-performance implementation of CFD largely depends on precision analog signal processing, which is not technology-friendly to ETROC in which 65 nm CMOS process and 1.2 V power supply are used.

Another way to correct time walk is to measure the signal size of each hit and then remove the dependence on it from the leading-edge (LE) time. The signal size could be obtained by measuring the amplitude, charge or TOT of the signal. When correcting the time walk with the amplitude or the charge, a peak detector or an integrator can be used, respectively. The analog values are then digitized with an analog-to-digital converter (ADC) and the time walk correction is performed in digital domain. However, the peak detector, the integrator and the ADC are not feasible for ETROC due to the high repetition rate of 40 MHz (LHC clock) and tight power budget. Containing the information of the signal size, TOT is more natural to be used in ETROC as it can be measured from the output pulse of the discriminator with a time-to-digital converter (TDC). What's more, TDC for leading edge measurement and TOT measurement can be merged together to save power consumption. Such a TDC has been designed with very low power consumption\cite{Wei_TDC2020} and is integrated with the analog front-end in the second prototype chip (ETROC1). A concern of time walk correction with TOT is that the jitter of trailing edge would contribute to the final time resolution, and this will be explained in this section. Table \ref{table:1} summarizes the time walk correction techniques discussed above. ETROC chooses to correct time walk with TOT due to its power efficiency and simplicity in implementation with the 65 nm process.

\begin{figure}[t]
\centerline{\includegraphics[width=3.5in]{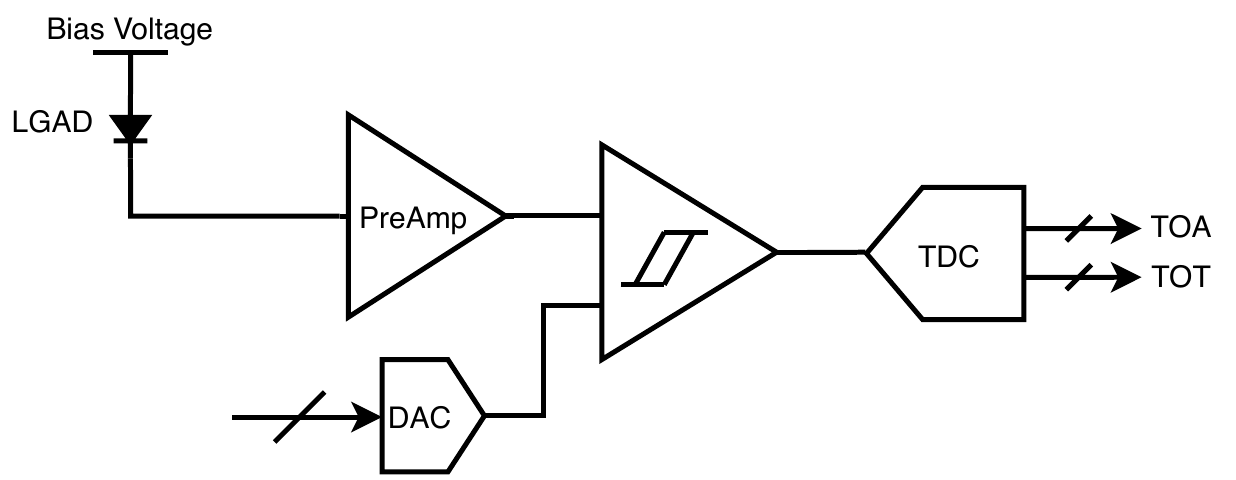}}
\caption{Design concept of the ETROC front-end circuits.}
\label{fig:fe}
\end{figure}

\begin{figure}[t]
\centerline{\includegraphics[width=3.5in]{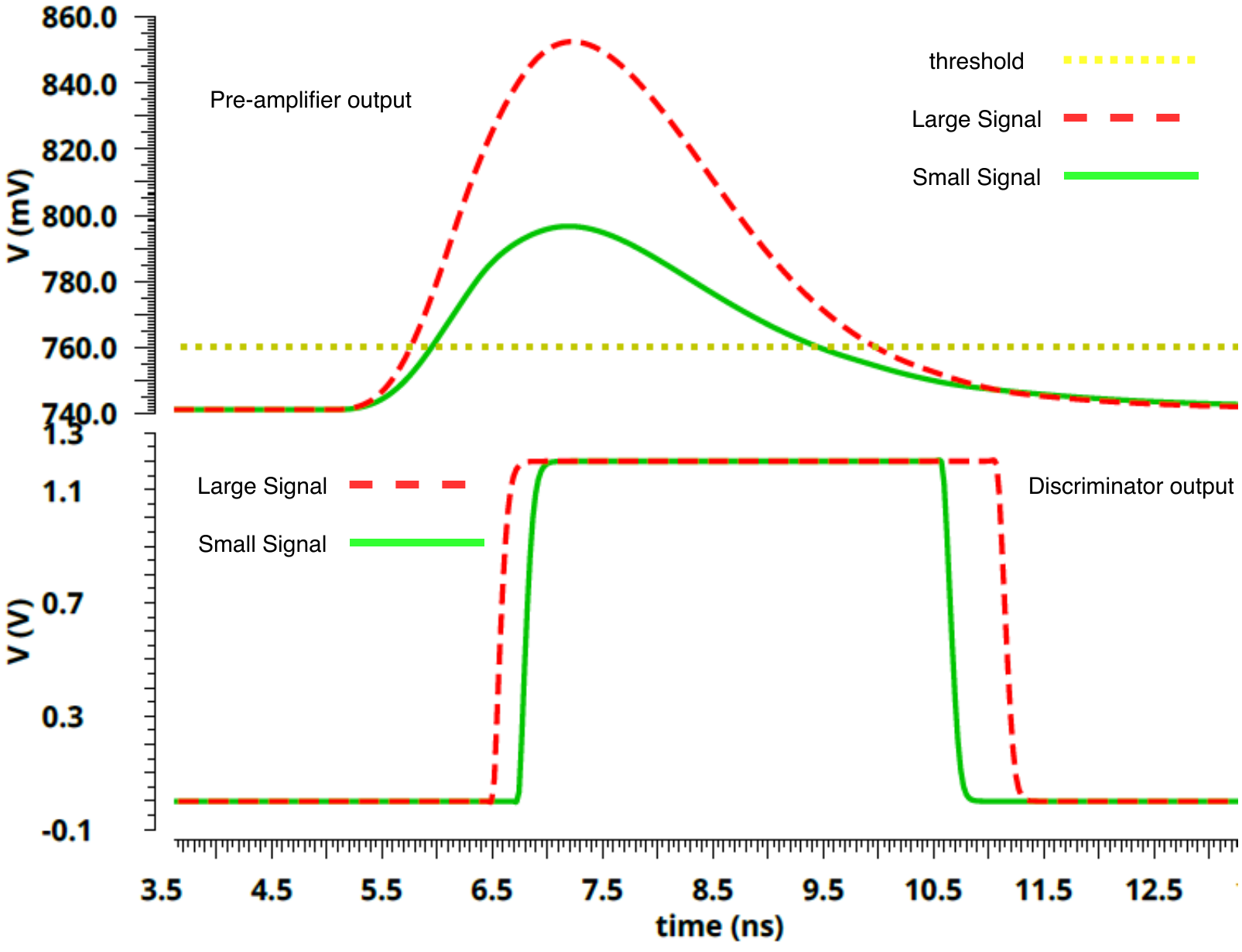}}
\caption{Conceptual waveform of the preamplifier and the discriminator.}
\label{fig:afe_wav}
\end{figure}

\figurename~\ref{fig:fe} shows the design concept of the ETL front-end circuits. The preamplifier connects to the LGAD with DC coupling, providing gain and bias to the LGAD sensor. The discriminator makes time discrimination at a given threshold voltage supplied by a digital-to-analog converter (DAC). The TDC measures TOA and TOT. The LGAD sensor being connected to ETROC with bump-bonding is reversely biased between a negative high voltage and the preamplifier input for setting up a proper intrinsic gain. 

\figurename~\ref{fig:afe_wav} illustrates a conceptual waveform of the preamplifier and the discriminator, with a small signal and a large signal. The leading edge crosses the threshold earlier while the trailing edge crosses the threshold later in the larger signal case, showing a larger time over threshold for a larger signal. This is valid for a properly designed timing measurement circuitry and can be used to correct time walk.

\begin{figure}[t]
\centerline{\includegraphics[width=3.1in]{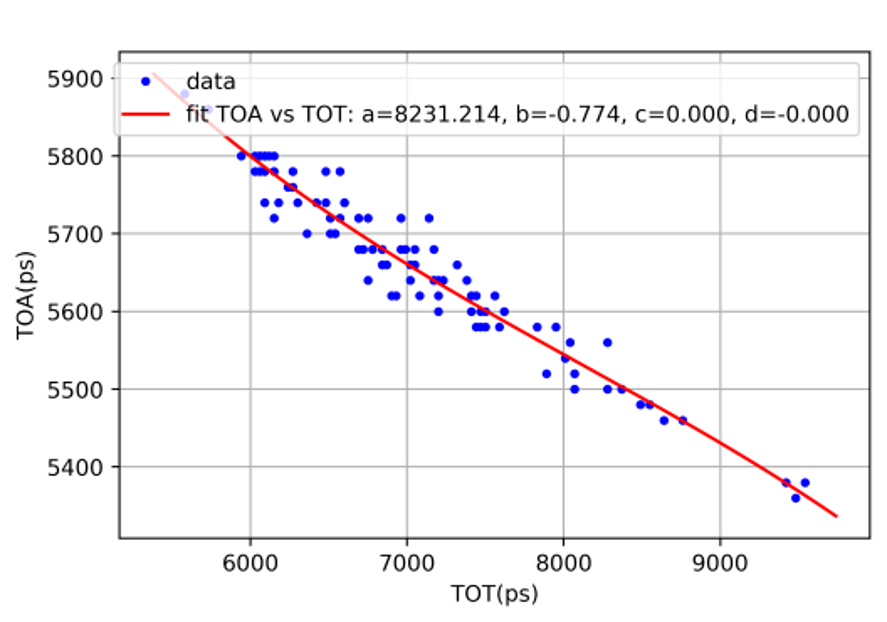}}
\caption{An example of TOA vs TOT plot and the curve of the fitted polynomial.}
\label{fig:toavstot}
\end{figure}

The TOA measured by the front-end shown in \figurename~\ref{fig:fe} is the digital representative of the leading-edge time including the time walk, which is correlated with the TOT, the digital representative of the time over threshold. \figurename~\ref{fig:toavstot} illustrates this dependence with a TOA vs TOT plot, using data from LGAD simulation as input to the analog front-end design. Data on the plot is fitted with a 3rd-order polynomial,
\begin{equation}
TOA_{fitted}=a+b\cdot~TOT+c\cdot~TOT^2+d\cdot~TOT^3
\label{eqe:7}
\end{equation}
, where a, b and c are parameters to be determined by the fitting. The time walk corrected TOA can be expressed as
\begin{equation}
\begin{split}
   &TOA_{corrected}=TOA-TOA_{fitted}\\
   &\approx~t_{le}+\epsilon_{TOA}-(a+b\cdot(t_{te}-t_{le}+\epsilon_{TOT}))\\
   &=-a+(1+b)\cdot~t_{le}-b\cdot~t_{te}+\epsilon_{TOA}-b\cdot~\epsilon_{TOT}\\
\end{split}
\label{eqe:8}
\end{equation}
, where $t_{le}$, $t_{te}$, $\epsilon_{TOA}$ and $\epsilon_{TOT}$ are the leading-edge crossing time, the trailing-edge crossing time, the quantization error due to TOA digitization and the quantization error due to the TOT digitization, respectively.
There is a constant offset between the corrected TOA and the actual time
of arrival, which can be determined from calibrations at the system
level. The first-order coefficient b, is a constant which typically ranges between -0.3 and -0.8, depending on the threshold setting and the dynamic behavior
of the front-end circuits. Higher-order coefficients represent non-linearity
dependency of the front-end circuits, which are usually small and can be
ignored in simplified analysis. Equation \eqref{eqe:8} implies that both the leading edge jitter and trailing edge jitter contribute to the time resolution. Other contributors to time resolution from the front-end
circuits include the quantization noise of TOA and TOT, the jitter due to the noise of the discriminator (and the associated threshold voltage), and the residual of the time walk
after correction. All these contributors can be added together in quadrature
to get the overall time resolution, as shown in equation \eqref{eqe:9}.
\begin{equation}
 \begin{split}
 &\sigma_t^2=\\
 &\sigma_{Landau}^2+(1+b)^2\cdot~\sigma_{LE}^2\\
 &+b^2\cdot~\sigma_{TE}^2+\sigma_{TOA}^2+b^2\cdot~\sigma_{TOT}^2+\sigma_{Discri}^2+\sigma_{TWR}^2   
 \end{split}
\label{eqe:9}
\end{equation}

Table \ref{table:2} summarizes the decomposition of the time resolution suggested by \eqref{eqe:9}.

Equation \ref{eqe:9} and table \ref{table:2} reveal some insights and design
considerations for optimization of the front-end circuits for ETROC:
\begin{enumerate}
\item Landau noise from the LGAD limits the best achievable time resolution
at about 30 ps.
\item Both the leading edge and the trailing edge jitter should be
optimized, and jitter contribution from trailing edges could be dominating in some cases. Fast edges are desired.
\item TDC bin size for TOA measurement can be as large as 40 ps, without contributing much (less than 12 ps) to overall time resolution.
\item TDC bin size for TOT measurement can be as large as 60 ps, without contributing much to the overall time resolution. The contribution is suppressed by $b$.
\item Time walk correction process can minimize the impact due to nonlinearity of front-end circuits.
\end{enumerate}

\begin{table*}[t]
  \centering
  \label{table:1}
  \caption{Time walk correction techniques summary.}
  \begin{tabular}{|m{4cm} ||m{6.2cm}|m{6.2cm}|}
    \hline
 Time walk correction techniques &Pros.&Cons.\\
 \hline
 Waveform Sampling & Best time resolution & Power and data bandwidth hungry\\
 \hline
 CFD   & Good performance    &Highly depending on analog signal processing\\
 \hline
 LE + amplitudes&   Natural for energy measurement applications  & Peak detector and ADC power consumption at high repetition rate\\
 \hline
 LE + charge &Natural for energy measurement applications, good linearity & Integrator and ADC power consumption at high repetition rate\\
 \hline
 LE + TOT    &Natural for time measurement, easy to scale down with technology and voltage & Trailing edge jitter sensitive\\
 \hline
  \end{tabular}
\end{table*}

\begin{table*}[t]
  \centering
  \label{table:2}
  \caption{Decomposition of time resolution.}
  \begin{tabular}{|m{4.5cm} ||m{1.2cm}|m{5.5cm}|}
    \hline
 Decomposition &Symbol&Comments\\
 \hline
 Landau noise   & $\sigma_{Landau}$    &30 ps for the baseline LGAD\\
 \hline
 Leading edge jitter&   $\sigma_{LE}$  & to be scaled by $1+b$\\
 \hline
 Trailing edge jitter & $\sigma_{TE}$ & to be scaled by $b$\\
 \hline
 TOA quantization noise    &$\sigma_{TOA}$ & $BinSize_{TOA}/\sqrt{12}$\\
 \hline
 TOT quantization noise    &$\sigma_{TOT}$ & $BinSize_{TOT}/\sqrt{12}$, to be scaled by $b$\\
 \hline
 Discriminator noise contribution    &$\sigma_{Discri}$ & Negligible with a proper design\\
 \hline
 Time walk residual after correction    &$\sigma_{TWR}$ & Negligible after correction, higher order polynomial helps to correct nonlinearity\\
 \hline
  \end{tabular}
\end{table*}

\begin{figure}[t]
\centerline{\includegraphics[width=3.1in]{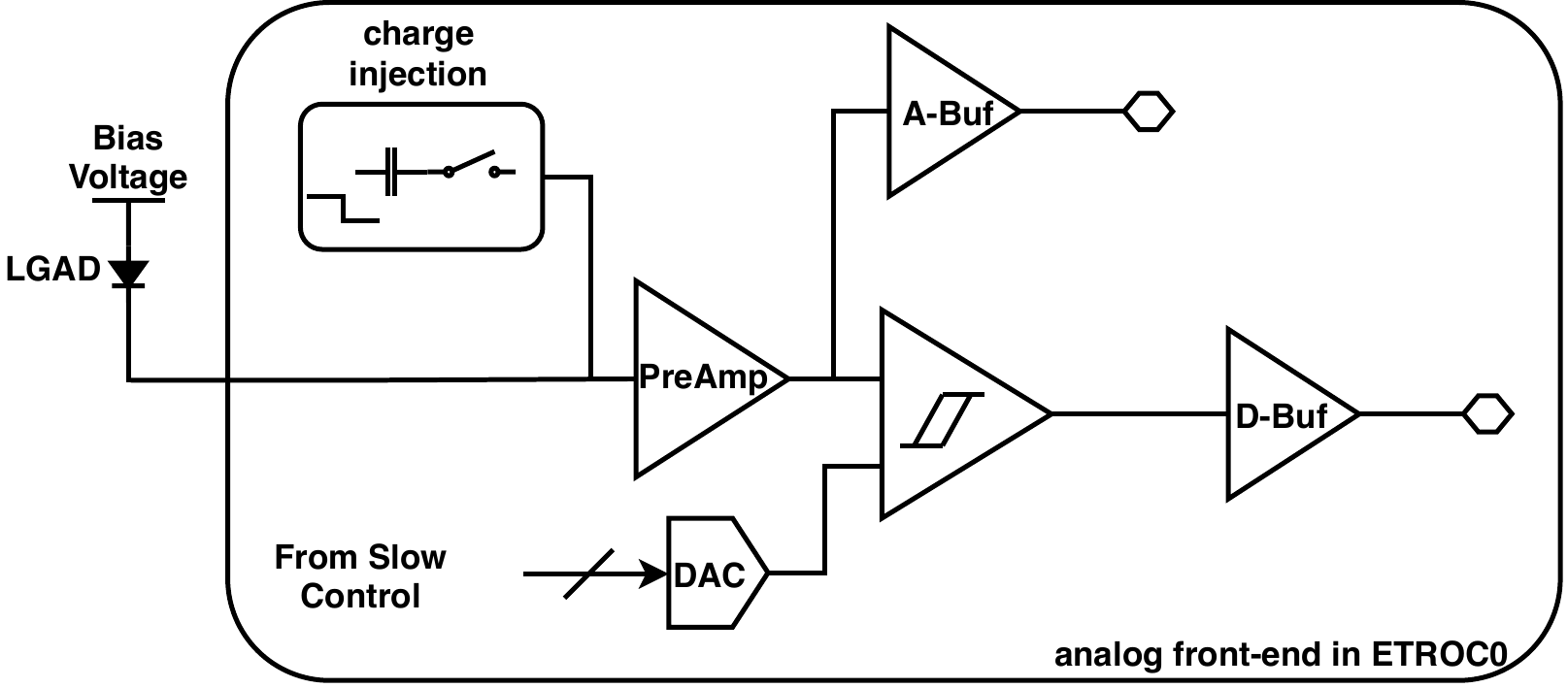}}
\caption{Block diagram of the analog front-end implemented in ETROC0.}
\label{fig:etroc0}
\end{figure}

\section{ETROC0 implementation}
Based on the design considerations described above, the first prototype chip for CMS ETL, ETROC0, has been developed, aiming for
exploring the performance of the analog part of the front-end readout
circuits (analog front-end). A preamplifier, a discriminator and a DAC for
threshold generation were implemented in ETROC0 to form the analog
front-end, as shown in \figurename~\ref{fig:etroc0}. The front-end performance evaluation and optimization has been done extensively using detailed LGAD simulation as input with post-layout simulation. A charge injector was
included to facilitate the test and calibration. Signal at the output of the preamplifier
and the discriminator are accessible through an analog buffer and a digital
buffer. The extra spying buffers could degrade the performance of the signal processing chain while driving heavy off-chip loads, and are designed to minimize the effect. 

\begin{figure}[t]
\centerline{\includegraphics[width=3.5in]{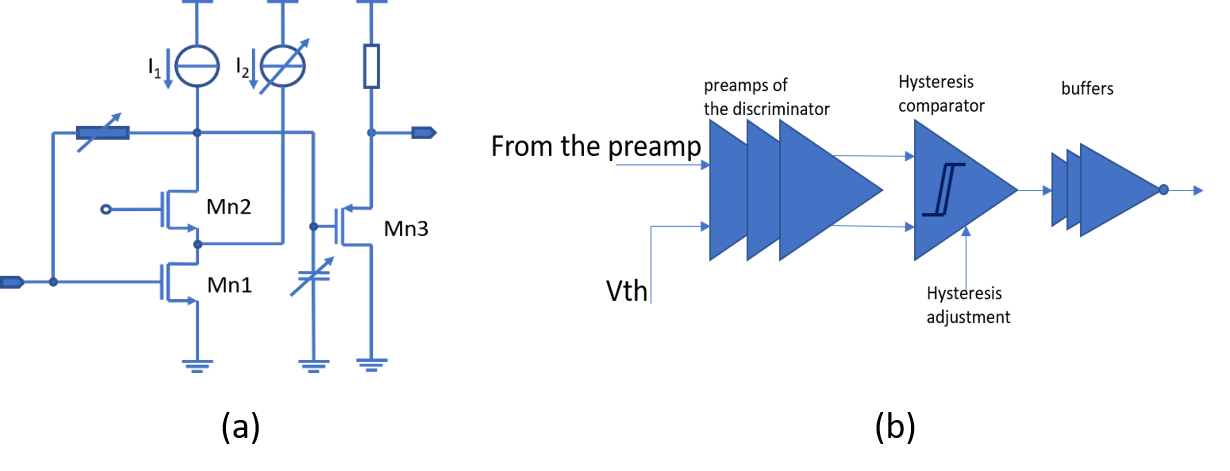}}
\caption{(a)simplified schematic of the preamplifier, (b) structure of the discriminator.}
\label{fig:pa_discri}
\end{figure}

\subsection{Preamplifier}
The ETROC0 preamplifier, as shown in \figurename~\ref{fig:pa_discri}(a), is a
buffered transimpedance amplifier (TIA). The size of the preamplifier is 90 x 94 ${\mu}m^2$. The core TIA is built with a
NMOS-input cascode amplifier and a feedback resistor. The feedback
resistance has four programmable settings between 4.4 and 20 kOhm, providing the flexibility of adjustment of gain and fall time. For example, at the beginning of the ETL operation, when LGAD has an adequate gain, a low feedback resistance could be used to obtain
fast trailing edge for optimum timing performance. As the gain of LGAD degrades due to irradiation,  a larger feedback resistance could be used to provide extra gain. Based on the simulation with LGAD data, a default feedback resistor of 5.7 $k\Omega$ is chosen. The load of the main brunch is a
PMOS cascode current source providing a constant current, $I_1$, which constrains the
slew rate of rising edge. A bias brunch provides additional current, $I_2$, to
the input transistor, $M_{n1}$, to boost its transconductance. The additional
bias current, $I_2$, is programmable with four settings to allow different trade-offs between power consumption and performance. The smallest bias current is the default setting. A better performance is to be expected at a larger bias
current. A programmable load capacitor at the output of the TIA provides
possibility to fine tune the rise time and bandwidth. The default setting is to use minimal additional load capacitance. A PMOS source
follower serves as a buffer and a level shifter at the same time.

\begin{figure}[t]
\centerline{\includegraphics[width=3.5in]{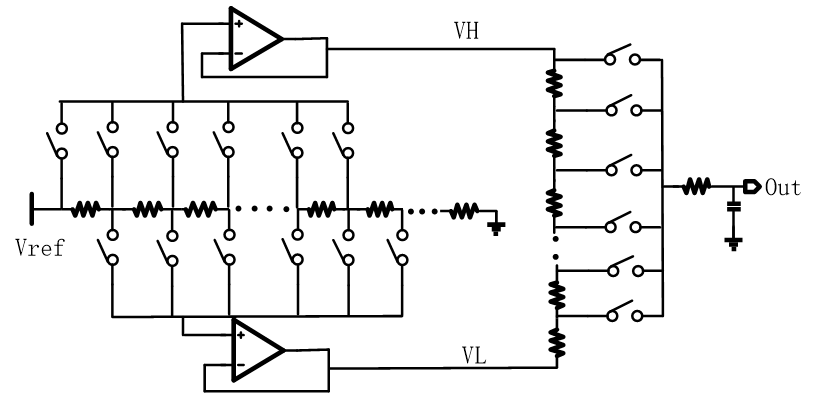}}
\caption{Simplified schematic of the 10-bit DAC.}
\label{fig:dac}
\end{figure}

\subsection{Discriminator and DAC}
The discriminator consists of a three-stage preamplifier, a comparator with programmable hysteresis and a buffer, as shown in \figurename~\ref{fig:pa_discri}(b). The size of discriminator is 81 x 67 ${\mu}m^2$. The input pulse seen by the discriminator could be very low in some extreme cases. The three-stage preamplifier in the discriminator is employed to amplify these small input pulses to a level where the succeeding comparator can fire. The comparator digitizes the differential input at the crossing point with an adjustable hysteresis ranging from 0 to 1 mV. Finally, the internal buffer is used to relieve the loading pressure from the following circuits.

One of the inputs of the discriminator is connected to the 10-bit DAC for in-pixel threshold generation. \figurename~\ref{fig:dac} shows the simplified schematic of the DAC. As the baseline of the preamplifier varies with temperature and bias setting, the DAC is designed to cover a range from 0.6 V to 1 V with a step size of 0.4mV. The DAC employs a resistor-string structure to achieve monotone over its operation range. A noise filter is used to limit the noise of the threshold voltage below 0.1 mV, to minimize the DAC noise contribution to the timing performance.

\subsection{Performance from Simulation}
Table \Romannum{3} summarizes the static parameters of the preamplifier obtained from simulation. The preamplifier consumes about twofold power at high-current bias than at low-current bias in order to achieve larger $dV/dt$ therefore a better timing performance. The output noise at high power is larger due to a larger bandwidth, the same is true at lower temperature. 
\begin{table}[t]
  \centering
  \label{table:3}
  \caption{The preamplifier static performance summary.}
  \begin{tabular}{|m{2cm} ||m{2.6cm}|m{2.6cm}|}
    \hline
 Parameter &Low power (@$-20^oC/25^oC$)&High power (@$-20^oC/25^oC$)\\
 \hline
 Power Consumption (mW)   & 0.70/0.76    &1.55/1.59\\
 \hline
 ENC (ke-)&   1.39/1.41  & 1.18/1.35\\
 \hline
 Gain (mV/fC) &2.6/2.1 & 3.1/2.7\\
 \hline
 Output-referred Noise ($mV_{rms}$)    &0.58/0.49 & 0.62/0.59\\
 \hline
 Input Resistance ($\Omega$)    &203/283 & 141/181\\
 \hline
  \end{tabular}
\end{table}

\begin{figure}[t]
\centerline{\includegraphics[width=3.5in]{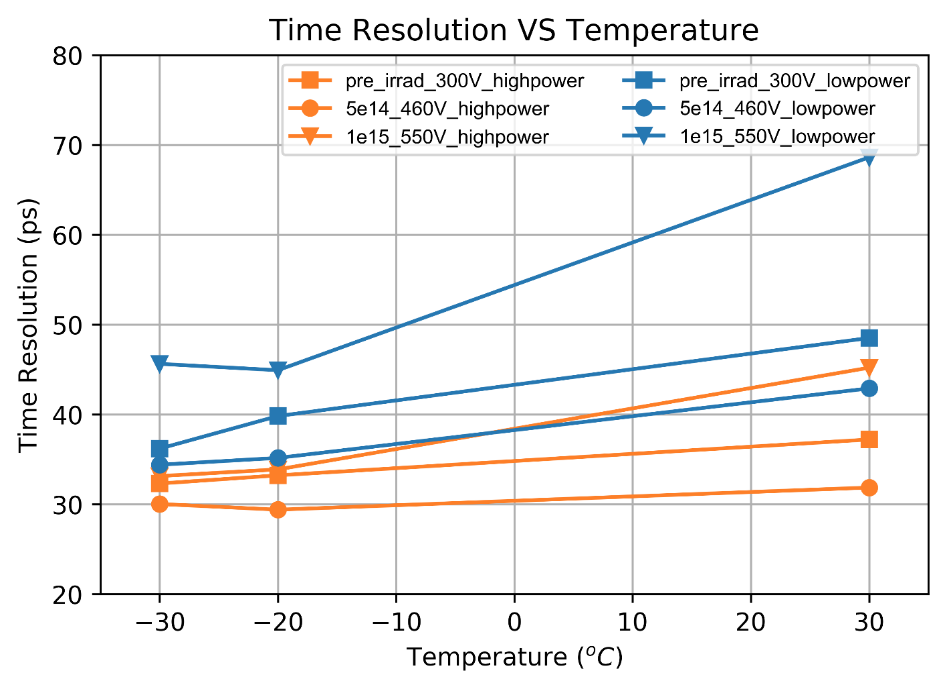}}
\caption{Analog front-end simulation with LGAD data from WeightField2.}
\label{fig:afesim}
\end{figure}
In order to evaluate and optimize the overall performance of the analog front-end in ETROC0 together with the LGAD sensor at the design stage, LGAD data from WeightField2 simulation\cite{CENNA2015149} was used in 
the circuits SPICE simulation at post-layout stage. LGAD data with different
irradiated levels were used to evaluate the overall timing performance, under different bias settings and temperature conditions. The results of the default power setting and a higher bias current setting (doubled) are shown in \figurename~\ref{fig:afesim}. With the default preamplifier bias setting, a 35 ps time resolution is obtained at the target operation temperature of $-20^oC$ when the module has been irradiated to $5\times10^{14}n_{eq}/cm^2$.

At the high fluence, the time resolution degrades to about 45ps because of reduced sensor gain. Better time resolution can be achieved with higher power consumption. \figurename~\ref{fig:afesim} shows the time resolution with the higher power preamplifier setting (doubling the power). At the operation temperature of $-20^oC$, at the higher power setting the time resolution is improved from the 35-45 ps to about 30-35 ps.  The overall temperature dependence has been improved a lot as well, especially for the high fluence case. 

\section{Test and Discussion}
ETROC0 is a 1 x 1 $mm^2$ chip in 65 nm CMOS process with 1.2 V power supply. Besides the analog
front-end circuits discussed, it includes a few other test structures and a
Serial Peripheral Interface (SPI) slave for the slow control. All individual design blocks can be tested separately, and the power consumption for each section can be measured separately as well.

ETROC0 bare die is glued onto a printed circuits board (PCB) and directly wire-bonded to the pads on the boards. Two sets of PCB have been developed to carry out the tests. One set is used for the charge injection and the TID test and the other set for the tests together with LGAD sensors.  

\begin{figure}[t]
\centerline{\includegraphics[width=3.5in]{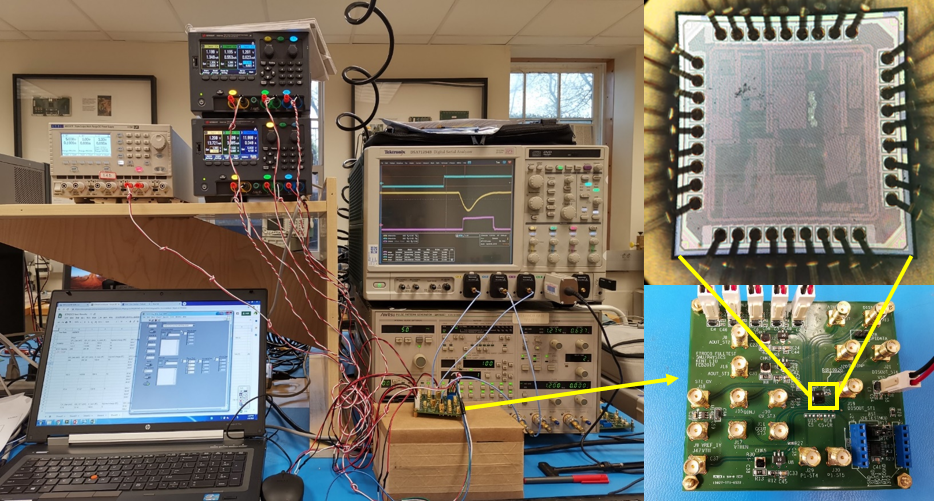}}
\caption{Setup of the charge injection test.}
\label{fig:qinjsetup}
\end{figure}
\subsection{Charge Injection Test}
ETROC0 has been extensively tested with charge injection through a 100 fF on-chip capacitor. The purpose of the charge injection test is to characterize the analog front-end without an LGAD sensor and carefully compare the measured results with the simulation. \figurename~\ref{fig:qinjsetup} shows the setup for the charge injection test and a photo of ETROC0. A 2 pF tantalum capacitor soldered on the PCB together with the 1.5 pF parasitic capacitance from the PCB is used to emulate equivalent capacitance of an LGAD sensor.

\begin{figure}[t]
\centerline{\includegraphics[width=3.5in]{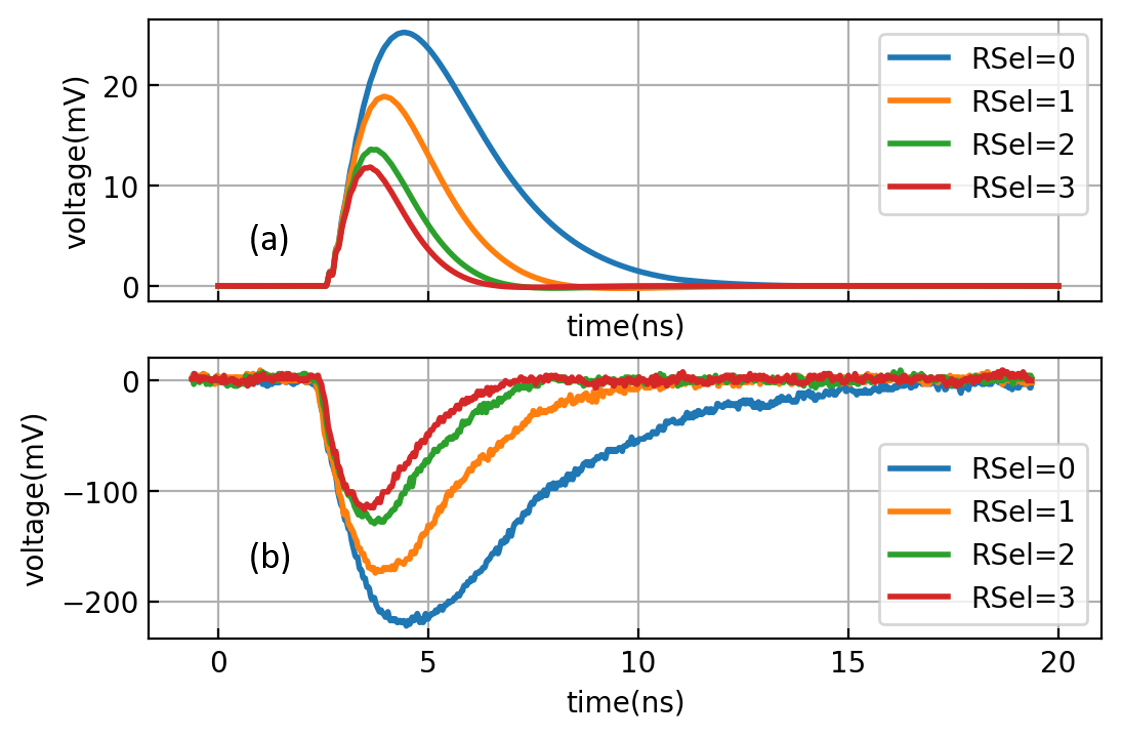}}
\caption{Waveform of the preamplifier with 6 fC charge input at different feedback settings, (a)simulated waveforms from the output of the preamplifier, (b)measured waveform from the on-board amplifier.}
\label{fig:pawav}
\end{figure}

Functionality of the preamplifier was examined by comparing the measured waveforms with the simulated. A commercial wide-bandwidth amplifier, Gali-S66+ with the gain of about -10, used on the board to amplify the signal before sending to an oscilloscope. The on-chip analog buffer following the preamplifier has a typical gain of 0.7 when driving the 50 $\Omega$ resistive load from the on-board amplifier. \figurename~\ref{fig:pawav} shows the simulated and the measured waveforms with 6 fC charge injected at different gain settings controlled by the feedback resistor. The agreement between the measured and simulated is good, within the expectation from the process variation.

\begin{figure}[t]
\centerline{\includegraphics[width=3.5in]{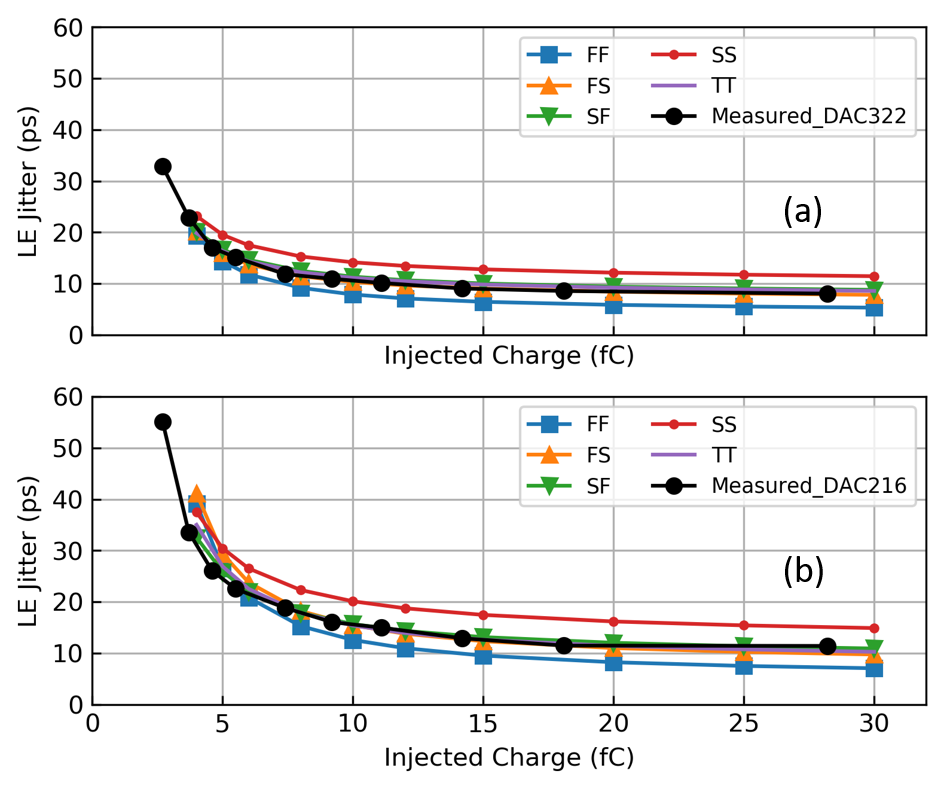}}
\caption{Leading edge jitter from measurement and simulation in various process corners(TT, FF, SS, SF, FS), (a)high power, (b)low power.}
\label{fig:afelejitter}
\end{figure}

The jitter performance of the full-chain front-end circuits was measured from the leading edge of the discriminator output and compared with simulation results. Taking advantage of the differential output of the pulse generator, one of the differential signals was used to inject charge, and the complementary signal was used as the trigger of the scope. This setup minimizes the jitter contribution from the pulse generator. The threshold was placed 9 DAC least significant bits (LSB), about 3.6 mV, above the measured baseline. \figurename~\ref{fig:afelejitter} shows the measured leading-edge jitter and the simulated at various process corners, at the lowest bias current and highest bias current. Since the charge injection pulse from the signal generator is fast (about 25 ps according to the specification), the signal seen by the preamplifier is faster than an actual LGAD signal. Nevertheless, the faster charge injection signal is an effective way to evaluate the performance of the analog front end (as if with an ideal sensor). The measured performance agrees well with the simulation. 
\begin{figure}[t]
\centerline{\includegraphics[width=3.5in]{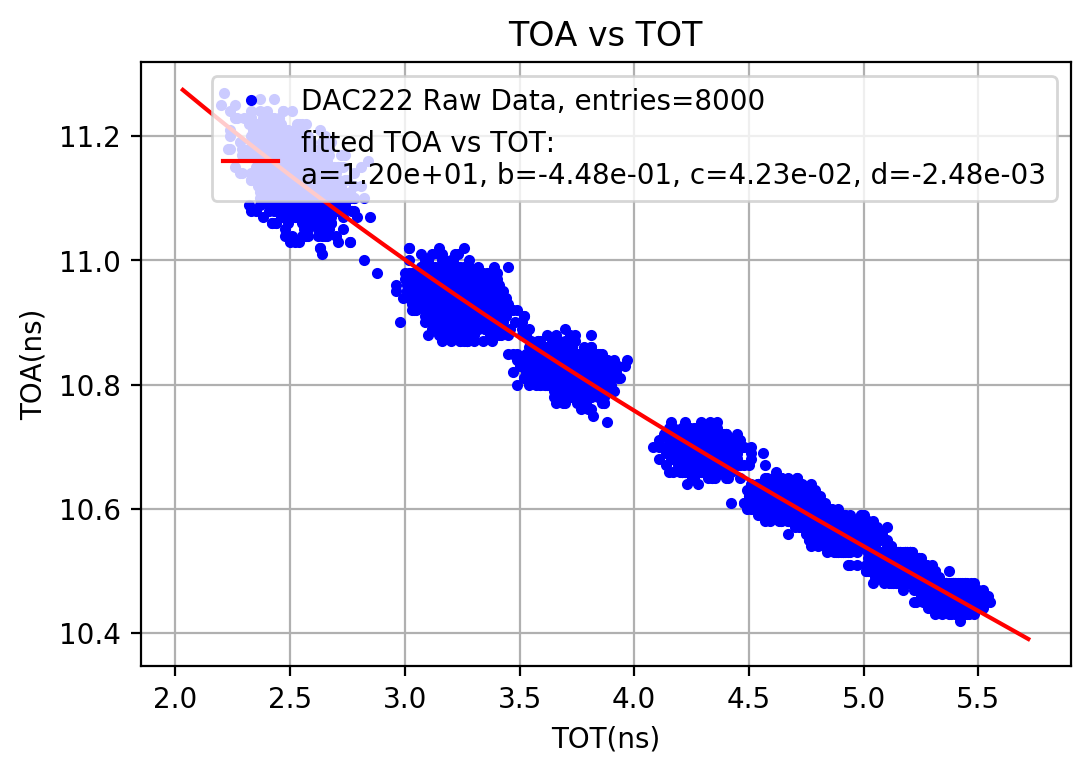}}
\caption{TOA vs TOT from charge injection test.}
\label{fig:totvstoa_qinj}
\end{figure}

The TOA vs TOT plot from charge injection test shown in \figurename~\ref{fig:totvstoa_qinj} demonstrates a health behavior of the analog front-end which is essential for time work correction with TOT.

The power consumption of the preamplifier is measured as 0.72 mW and 1.51 mW at low-power and high-power bias, respectively. The discriminator power consumption is measured as 0.84 mV. All the measured power consumptions match well with simulation. 

\begin{figure}[t]
\centerline{\includegraphics[width=3.5in]{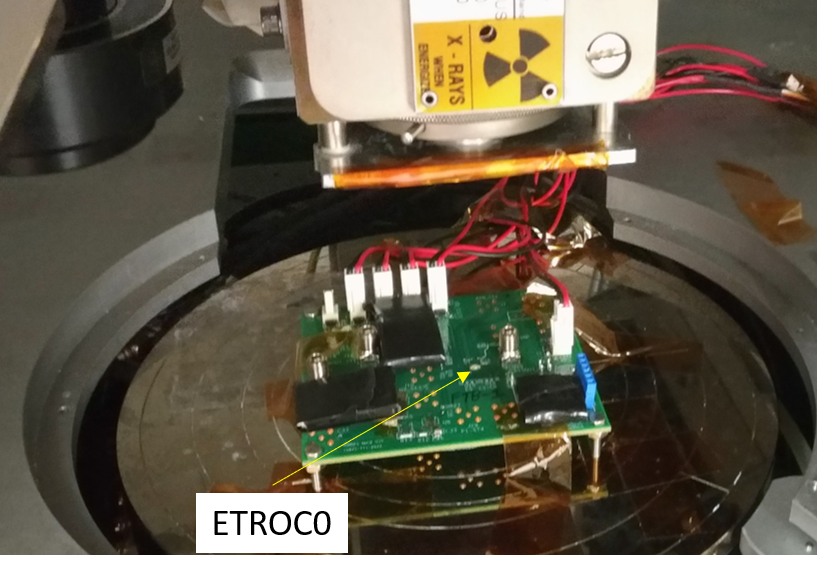}}
\caption{TID test setup at CERN X-ray irradiation facility (AsteriX).}
\label{fig:tidsetup}
\end{figure}

\begin{figure}[t]
\centerline{\includegraphics[width=3.5in]{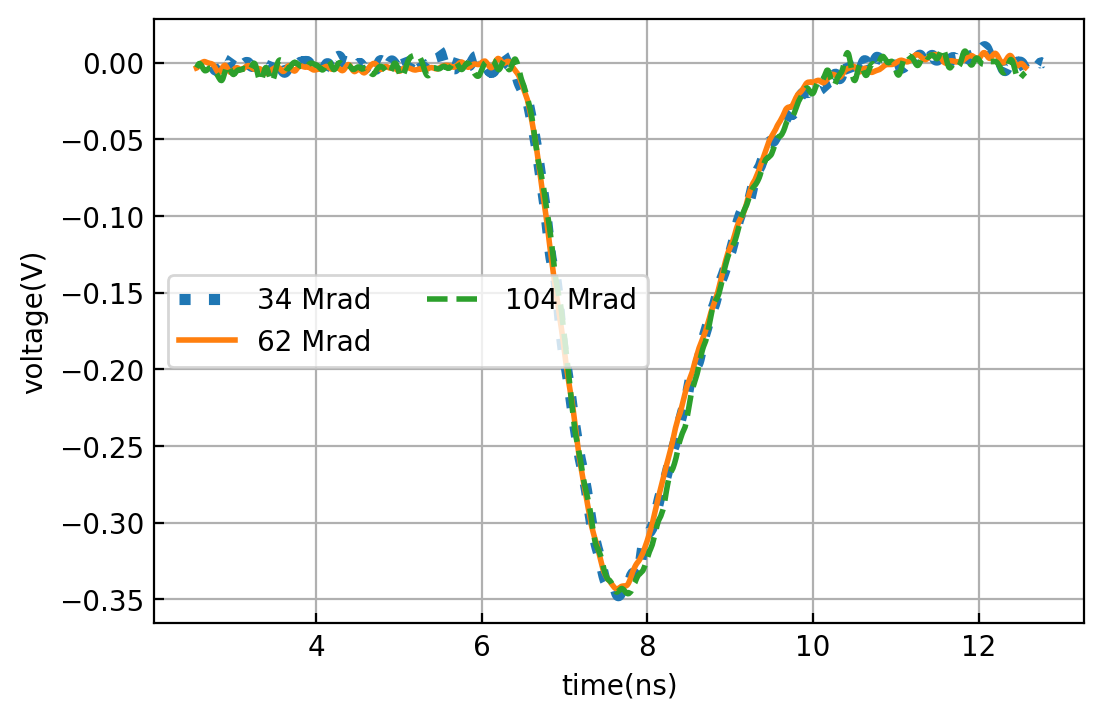}}
\caption{The preamplifier waveform during the irradiation test.}
\label{fig:tidwav}
\end{figure}
\subsection{TID Test}
A few ETROC0 chips were irradiated to 100 Mrad (ETROC spec.) with x-ray for TID test at CERN. \figurename~\ref{fig:tidsetup} shows the test setup at CERN. The x-ray generator is AsteriX with tube voltage of 40 kV and tube current of up to 50 mA. The calibrated dose rate is up to 8.09 Mrad/hr. ETROC0 was directly exposed to x-ray while other components on the board were shielded with 6-mm aluminum covers. For the TID testing, no input capacitor was loaded on the ETROC0 boards to emulate the LGAD capacitance. Without the extra input capacitance, the test will be more sensitive to any change due to irradiation. 

The current of the preamplifier and the discriminator were monitored during the test. The preamplifier current was found to drop by 5\% and the discriminator current was found to drop by 10\% during irradiation. Waveform of the preamplifier using charge injection was recorded at some dose points during the irradiation and no difference was observed, as shown in \figurename~\ref{fig:tidwav}. The chips were then extensively tested after the irradiation over a few weeks and no performance degradation was observed when compared with the performance before the TID testing (as shown in \figurename~\ref{fig:afelejitter}). 
\begin{figure}[t]
\centerline{\includegraphics[width=3.5in]{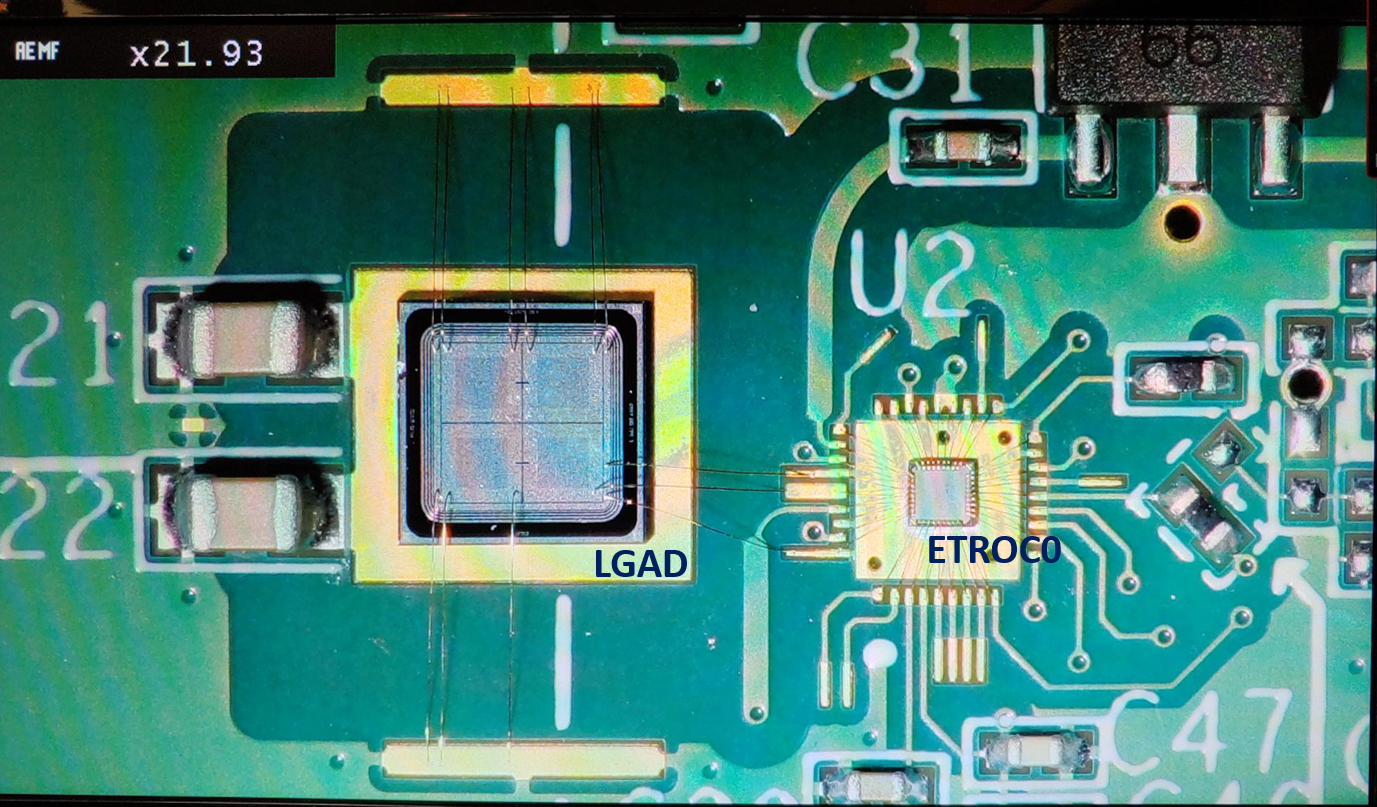}}
\caption{LGAD-ETROC0 board. The LGAD used were 2x2 sensor with only one pixel wire bonded to the ETROC0.}
\label{fig:etroc0lgad}
\end{figure}

\begin{figure}[t]
\centerline{\includegraphics[width=3.5in]{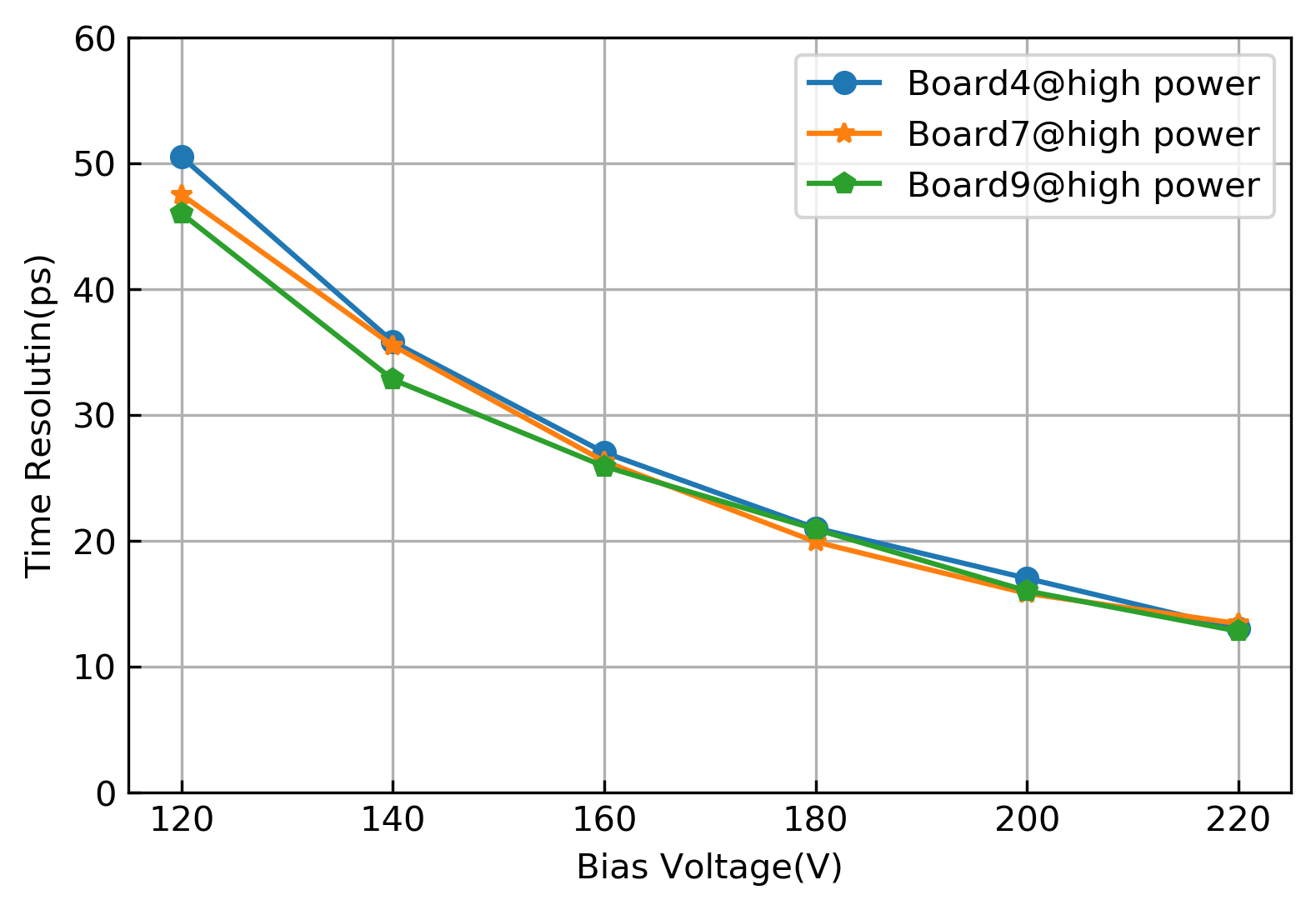}}
\caption{Time resolution vs HV scan using laser for a few ETROC0 boards using HPK LGAD sensor, to prepare for the beam test.}
\label{fig:laser}
\end{figure}

\begin{figure}[t]
\centerline{\includegraphics[width=3.5in]{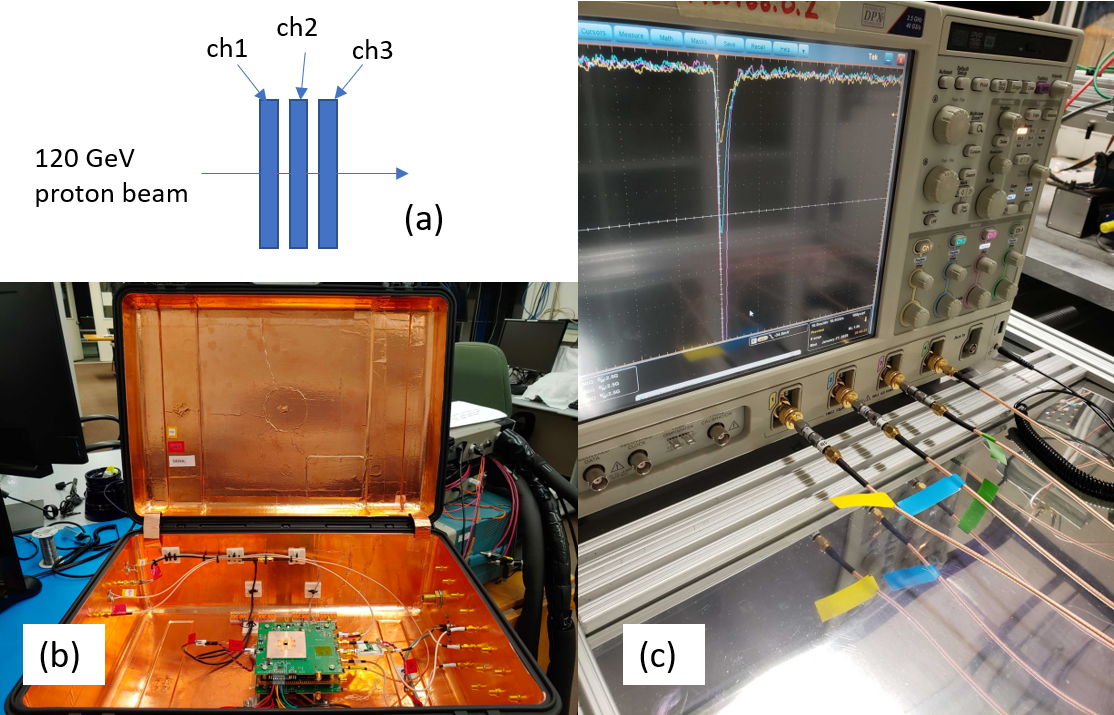}}
\caption{A simple telescope with 3 HPK sensor based ETROC0 boards for ETROC0 beam test at Fermilab with 120 GeV proton beam.}
\label{fig:beam_setup}
\end{figure}

\begin{figure}[t]
\centerline{\includegraphics[width=3.5in]{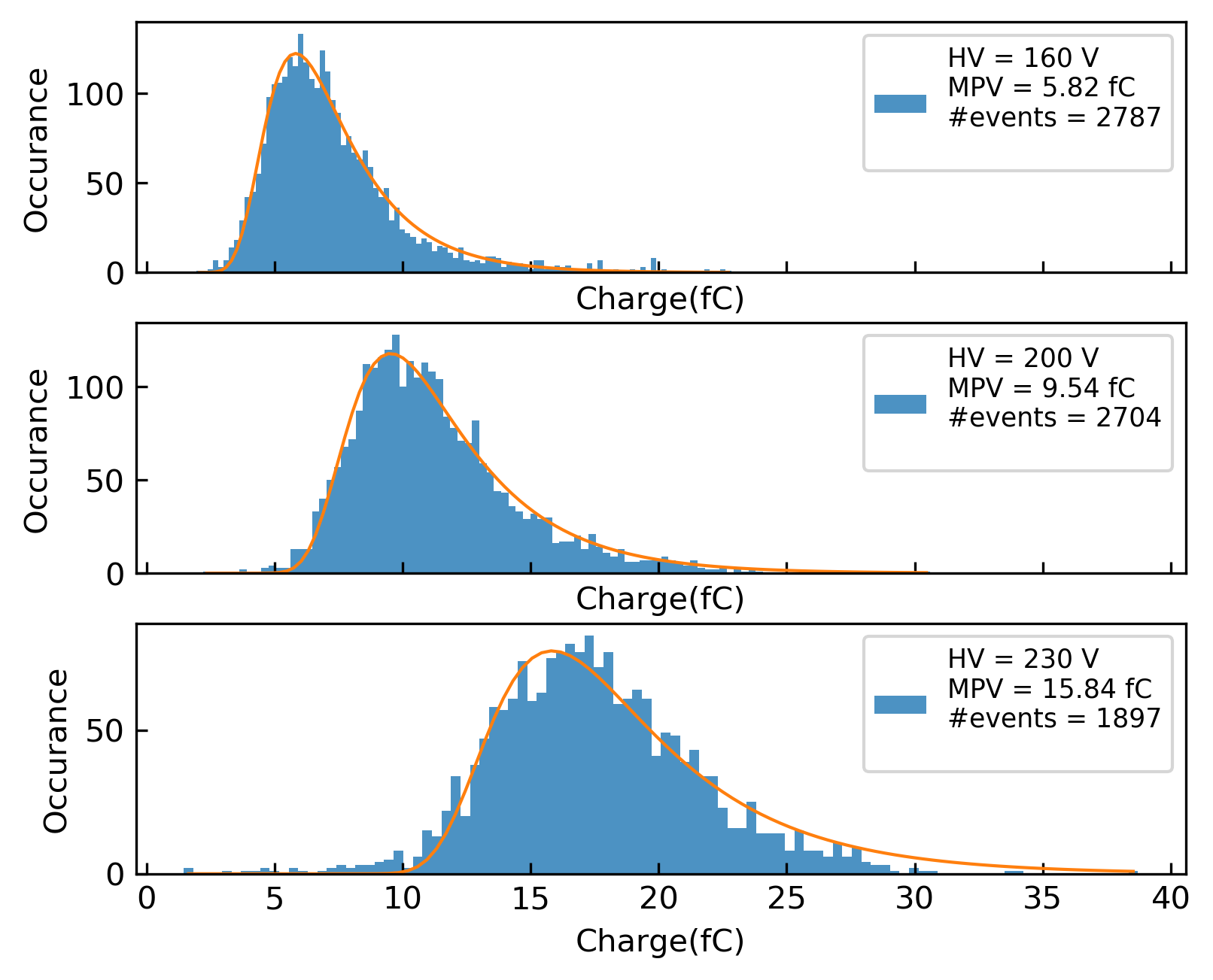}}
\caption{Estimated charge distribution of the BUT with different sensor bias voltage.}
\label{fig:chargehist}
\end{figure}

\begin{figure}[t]
\centerline{\includegraphics[width=3.5in]{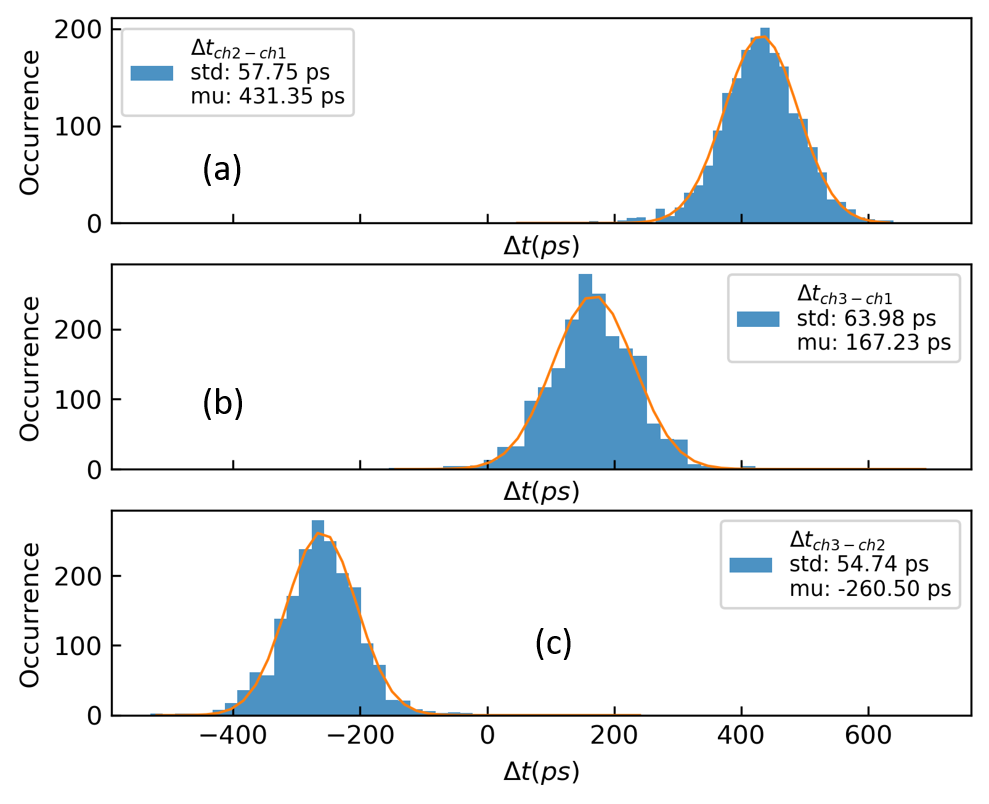}}
\caption{Inter-channel TOA of the preamplifier waveform with CFD.}
\label{fig:beamhist}
\end{figure}

\begin{figure}[t]
\centerline{\includegraphics[width=3.5in]{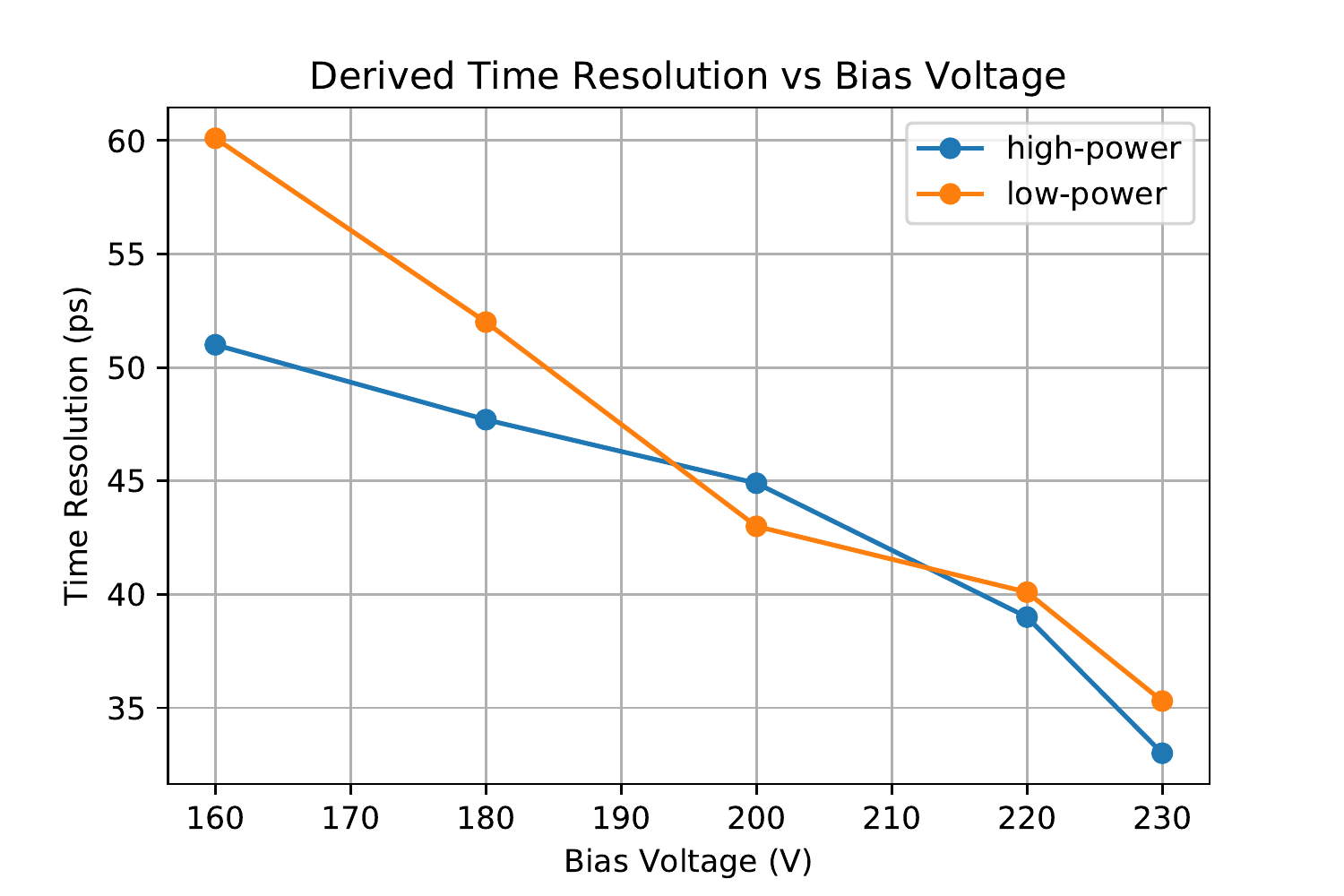}}
\caption{Time resolution of the preamplifier at different bias voltages.}
\label{fig:tres_scanhv}
\end{figure}

\subsection{Beam Test}
\subsubsection{Beam Test Preparation}
To prepare for the beam test, the LGAD-ETROC0 boards were first tested using a 1053 nm laser head driven by an Advanced Laser Diode Systems EIG2000DX laser controller. The LGAD used were 2x2 sensor with only one pixel wire bonded to the ETROC0, as shown in \figurename~\ref{fig:etroc0lgad}. The laser was positioned with a Zaber T-LSM XYZ stack and focused with a Schafter+Kirchhoff collimator and lens. Good consistency is observed among different boards. \figurename~\ref{fig:laser} shows the time resolution measured with laser for three ETROC0 boards using HPK LGAD sensors. A few other ETROC0 boards used FBK LGAD sensors, similar consistency is also observed among those boards. 

To test the LGAD-ETROC0 response to charge particles while waiting for beam, and to study the long term operational robustness of the LGAD-ETROC0 combination, a simple cosmic ray telescope was formed by a ETROC0 board with FBK LGAD sensor and a silicon photomultiplier (SiPM) with Lutetium–yttrium oxyorthosilicate (LYSO). The LYSO-SiPM serves to confirm the cosmic ray with a proper amplitude threshold. The cosmic ray telescope operated continuously over a period of two months, and the LGAD-ETROC power supplies (current and voltage) as well as the trigger rate were monitored. The ETROC0 telescope worked continuously without problem over the two months period, and collected more than one thousand cosmic ray events. The events served as the initial study of the LGAD-ETROC0 full chain using charged particles before beam. 

\subsubsection{Beam Test with A Simple Telescope}
The first ETROC0 beam test was performed at the FermiLab test beam facility in early 2020 at room temperature, using a telescope formed by LGAD-ETROC0 boards. A simple telescope with three LGAD-ETROC0 boards was used for beam test with a 120 GeV proton beam.  \figurename~\ref{fig:beam_setup} shows the beam test setup, with the ETROC0 preamp signal waveforms displayed on the scope for all three channels during the beam test. The preamplifier waveforms of channel 1 and channel 3 are used as reference and the channel 2 is the board under test (BUT). The preamplifier waveform of each channel and the discriminator waveform of channel 2 are sampled at 10 GS/s in real-time with an oscilloscope and the data is then analyzed off-line. For this setup, the LGAD sensors used for channel 1 and channel 3 were FBK LGAD sensor, while channel 2 used HPK LGAD sensor. During the testing, the bias voltage of the FBK sensors on channel 1 and channel 3 were fixed (at -300V) while the bias voltage of the HPK sensor on channel 2 were scanned from -160 V to -230 V.

In this simple 3-layer telescope, no dedicated precision time reference is used, instead, each ETROC0 board is the time reference of the others. The inter-channel time resolution can be expressed as
\begin{equation}
 \left\{ \begin{array}{lll}
\sigma_{21}^2=\sigma_{2}^2+\sigma_{1}^2 , \\
\sigma_{31}^2=\sigma_{3}^2+\sigma_{1}^2 , \\
\sigma_{32}^2=\sigma_{3}^2+\sigma_{2}^2 . 
\end{array} \right. 
\label{eqe:10}
\end{equation}
The time resolution of the BUT is derived from \eqref{eqe:10} as:
\begin{equation}
\sigma_2=\sqrt{0.5\cdot(\sigma_{32}^2+\sigma_{21}^2-\sigma_{31}^2)}.
\label{eqe:11}
\end{equation}

The charge collected by the sensor can be estimated from the preamplifier waveform, with the integral of the preamplifier output voltage over time divided by the trans-impedance of the preamplifier. \figurename~\ref{fig:chargehist} shows the charge distribution at some bias voltages for channel 2. The simulated trans-impedance of 4.4 KOhm is used in charge estimation. A typical gain of 0.7 from the on-chip analog buffer and the gain of on-board amplifier are taken into account as well. 

\figurename~\ref{fig:beamhist} shows the inter-channel time differences for the three channels and the standard deviations of the distributions, with the channel 2 (the BUT) LGAD biased at -230V and preamplifier at high power mode.  The time of each channel is determined using the preamplifier waveform with CFD applied. The time resolution of the ETROC0 preamplifier on the BUT is derived as 33 ps from \eqref{eqe:11}. \figurename~\ref{fig:tres_scanhv} shows the time resolution derived from the pre-amplifier at different bias voltage for low-power and high-power settings. The ETROC0 time resolution derived from the discriminator output is about 41 ps, after TOT based time walk correction applied. Although the measured discriminator performance is still within specification, it is not as good as what the simulation suggested. As will be discussed next, this could have something to do with the fact that the signals were observed through extra spying buffers and will be discussed below.

\subsection{Discussion}
 It is worth pointing out that ETROC0 front-end performance is evaluated by spying on the preamplifier and discriminator output using extra analog and digital buffers. This approach has worked reasonably well for ETROC0, though the extra buffers could affect the performance of the actual signal processing chain and the measurements may not reflect the true performance of the analog front-end. Based on our experience with ETROC0 testing, the testing board needs to be carefully designed in order to minimize the disturbance of the delicate front-end signal processing circuitry. 
The ultimate way to evaluate the true performance of the front-end is to directly use the TDC to measure the timing, and this is what is done in the second prototype chip, ETROC1. ETROC1 has the TDC stage integrated with the preamplifier and discriminator, allowing the testing of the full chain without any spying of the preamp and discriminator output.

The initial beam test in early 2020 was interrupted in March 2020, and next round of beam test is scheduled for early 2021 at Fermilab during which the ETROC1 chips will be used to study the performance of the front-end without any spying of the preamplifier output and the discriminator output. The ETROC1 has exactly the same front-end design as ETROC0.

\section{Conclusions}
This paper presents the first analog front-end prototype (ETROC0) design for the CMS ETL readout chip, implemented in a 65 nm CMOS ASIC. The front-end design considerations have been presented to explain how the front-end design architecture was chosen for ETROC, this is followed by some of the key design features of the building blocks, including the preamplifier, discriminator and supporting circuitry. The design methodology of using the LGAD simulation data to evaluate and optimize the front-end design is emphasized. The ETROC0 prototype chips have been extensively tested using charge injection and the measured timing performance and power consumption agree well with simulation, and have been also tested successfully with TID up to 100Mrad. The preliminary beam test results are also very promising, more beam test results will be presented in the near future. The ETROC0 front-end design has been used in the next prototype chip, ETROC1, without modification.

\section{Acknowledgements}
The authors would like to acknowledge Nicolo Cartiglia (INFN Torino) for providing the LGAD simulation data and many useful discussions, Abe Sedan, Hartmut Sadrozinski, Zachary Galloway and Simone Mazza from the UCSC group for their help with test beam preparation, Paulo Moreira and Szymon Kulis(both from CERN) for providing the IOs in ETROC0, Christophe de LA TAILLE and Nathalie SEGUIN-MOREAU (both from OMEGA Ecole polyechnique) for useful discussions.

The authors also acknowledge the ETROC God Parents Committee for useful discussions and suggestions. The committee members include David Christian (FNAL), Gary Drake (FNAL), Carl Grace (LBNL), Christine HU-GUO (IPHC/IN2P3), Ron Lipton (FNAL), Eric Oberla (University of Chicago), Fukun Tang (University of Chicago) and Gary Varner (University of Hawaii).

\bibliography{bibtex/ref.bib}{}
\bibliographystyle{IEEEtran}

\end{document}